\documentclass[preprint,10pt]{elsarticle}

\usepackage{amssymb,amsmath}
\def\be{\begin{equation}}
\def\ee{\end{equation}}

\usepackage{graphicx,eucal}
\usepackage[all]{xy}
\xyoption{all}

\usepackage{color}

\journal{Physica A}

\begin{document}

\begin{frontmatter}
\title{A Fresh Look at the ``Hot Hand'' Paradox}
\author{S. Redner}
\address{Santa Fe Institute, 1399 Hyde Park Road, Santa Fe, New Mexico, 87501}
%\date{\today}

\begin{abstract}
  We use the backward Kolmogorov equation approach to understand the
  apparently paradoxical feature that the mean waiting time to encounter
  distinct fixed-length sequences of heads and tails upon repeated fair coin
  flips can be different.  For sequences of length 2, the mean time until the
  sequence HH (heads-heads) appears equals 6, while the waiting time for the
  sequence HT (heads-tails) equals 4.  We give complete results for the
  waiting times of sequences of lengths 3, 4, and 5; the extension to longer
  sequences is straightforward (albeit more tedious).  We also derive moment
  generating functions, from which any moment of the mean waiting time for
  specific sequences can be found.  Finally, we compute the mean waiting
  times $T_{2n\rm H}$ for $2n$ heads in a row, as well as the moment
  generating function for this sequence, and $T_{n\rm(HT)}$ for $n$
  alternating heads and tails.  For large $n$,
  $T_{2n\rm H}\sim 3 T_{n\rm(HT)}$.  Thus distinct sequences of coin flips of
  the same length can have very different mean waiting times.
\end{abstract}

\end{frontmatter}

% \maketitle

\section{Introduction}

This article is dedicated to Charlie Doering, who left this world much too
soon.  While I didn't have the opportunity to collaborate with him, I did
have the pleasure of many fruitful and engaging scientific and social
interactions with him over nearly four decades.  It was always enjoyable to
discuss science, and indeed almost any topic, with Charlie because of his
enthusiasm, his insights, and his ability to make you feel good about what
you were presenting to him.  He was a true mensch; the world would be a much
better place if he were still with us.

The account of the ``hot hand'' phenomenon that is presented here is a
``golden oldie'' that has been extensively investigated in both the serious
and the popular literature.  Even though this topic was not close to
Charlie's recent interests, I'm pretty sure that he would have liked this
story and that he'd have insightful suggestions that would have improved this
article.

In repeated flips of a fair coin, the outcomes H (heads) or T (tails)
occur equiprobably.  Thus in a long string of $N$ coin flips, the
number of heads and tails, $N_H$ and $N_T$, will be nearly equal, with
$|N_H-N_T|$ of the order of $\sqrt{N}$.  Given that H and T appear
equiprobably, the average frequencies of specific fixed-length
sequences of H's and T's must be the same; for example, the sequence
HHTH occurs with the same frequency as HTTH.  One might then expect
that the waiting time before encountering either of these sequences
should be the same.  Surprisingly, this expectation is false!  This
phenomenon is now known as the ``hot hand'' paradox or the ``hot
hand'' fallacy.

The counterintuitive nature of this ``hot hand'' paradox appears to
have first been studied in a systematic way by Gilovich et
al~\cite{GVT85}.  These authors sought to understand if scoring
streaks of individual players in basketball games was merely a
manifestation of random fluctuations or whether a player's scoring
could be characterized by well defined ``hot'' and ``cold'' streaks.
Their conclusion was that a ``hot hand'' did not exist in basketball
scoring statistics.

However, it was later found that there does exist a hot hand paradox
when the problem is formulated in an appropriate way. Namely, in a
long string of fair coin flips, the waiting times for the next
appearance of distinct sequences of equal-length strings of H's and
T's (for example, HHTH and HTTH), can be different.  This form of the
hot hand paradox is perhaps best stated in the following stark way (as
mentioned at the outset of the article by Miller and
Sanjurjo~\cite{MS18}): suppose one flips a fair coin $N$ times.
Whenever an H occurs (there should be roughly $\frac{1}{2}N$ such
H's), one then records the outcome of the next coin flip. Naively, one
expects that that number of recorded H's should be one-half of the
total number of H's, i.e., $\frac{1}{4}N$ such events.  In fact, this
number is less than $\frac{1}{4}N$\,!

The seemingly paradoxical nature of the ``hot hand'' phenomenon has
spawned considerable discussion and literature that has ultimately
resolved the paradox (see, e.g.,
Refs.~\cite{MS18,G88,BW91,RB92,RB97,JA10,RV10,MS19,RBG19}).  However,
the approaches given in some of these references are complicated and
the simplicity of the mechanism that underlies the paradox can be lost
in calculational details; a notable exception, however is
Ref.~\cite{A11}, which provides an extraordinarily simple and
appealing way to determine the mean waiting time for an arbitrary
sequence of heads and tails of arbitrary length.

Here we give an alternative route to understand the hot hand paradox that is
based on the backward Kolmogorov equation~\cite{K31,R01,BMS13}.  This
formulation has proved to be extremely useful in a variety of first-passage
phenomena.  By recasting the hot hand paradox as a first-passage problem, we
can use the backward Kolmogorov equation to compute the waiting time for
specific sequences of H's and T's of length up to 5.  This method can be
straightforwardly extended to longer sequences if so desired.  We also give
an intuitive reason why different sequences of the same length do not occur
with the same frequency.

We then extend the backward Kolmogorov approach to derive the moment
generating function for all sequences of length up to 3; again, this approach
could be readily extended to longer sequences if so desired.  From the
generating function, arbitrary-order moments of the waiting time can be
easily extracted.  These results seem to have not been derived previously.
Finally, we also compute the waiting time for particularly simple sequences
of arbitrary length, namely, the sequence of $2n$ consecutive H's and the
sequence of $n$ consecutive (HT)'s.  We find that
$T_{2n\rm H}\sim 3 T_{n\rm(HT)}$, so that $2n$ heads in a row is three times
less frequent than $n$ (HT)'s in a row.

To complete this introduction, we now present the simple idea that
underlies the backward Kolmogorov equation.  Consider a Markov process
that is currently in a particular state $S$.  We want to compute the
average time $T_{S\to F}$ until the process reaches a specified final
state $F$.  Suppose that there are two possible outcomes at each stage
of the process that occur with equal probability.  That is, from state
$S$, the process transitions either to state $S'$ or to $S''$, each
with probability $\frac{1}{2}$.  Suppose further that the time
required for each transition equals 1.  Since the Markov process has
no memory, when either of the states $S'$ or $S''$ are reached, the
process starts anew.  Consequently, the hitting time when starting
from $S$ is just the average of the hitting times starting from either
$S'$ or $S''$ plus the time spent in the transition itself.  That is
\begin{align}
  \label{K}
  T_{S\to F} =\tfrac{1}{2}(T_{S'\to F}+1) + \tfrac{1}{2}(T_{S''\to F}+1)\,.
\end{align}
We will use this basic equation to compute the waiting time for specific
sequences of H's and T's of a given length as a result of repeated flips of a
fair coin.

A powerful aspect of the backward Kolmogorov approach is that it applies to
\emph{any functional} of the first-passage time and not just the average
first-passage time.  Thus one can write equations similar to \eqref{K} for
the mean-square time, the mean-cube time, etc.  Even more striking, we can
write an equation of the form of \eqref{K} for the moment generating
function, $\langle e^{-st}\rangle$, where the angle brackets denote taking
the average over all possible sequences of heads and tails, from which
arbitrary-order moments can be obtained merely by a Taylor-series expansion.

\section{Average Waiting Times}

\subsection{Doublets}

Let us start with the simplest example of length-2 sequences.  The possible
sequences are HH, HT, TH, and HH.  Because the coin is fair, we obtain the
same statistics by the substitution $H\leftrightarrow T$, so that the waiting
time for the sequences TT and TH is the same as that for HH and HT.
Consequently, we only consider the first two sequences.  How long does one
have to wait before encountering each of these sequences in a long string of
fair coin flips?

Starting with Eq.~\eqref{K}, we first compute the mean waiting time
$T_{\rm HH}$ to encounter an HH sequence.  For this purpose, we introduce the
auxiliary restricted times: \vspace{-2mm}
\begin{itemize}
  \itemsep -0.5ex
\item $A$, the average waiting time for the sequence HH starting with an H.
\item $B$, the average waiting time for the sequence HH starting with a T.
\end{itemize}
These two times obey the backward equations
\begin{subequations}
\begin{align}
\begin{split}
  \label{AB1}
  A&= \tfrac{1}{2}\times 2 +\tfrac{1}{2}(1+B)\\
  B&= \tfrac{1}{2}(1+B) +\tfrac{1}{2}(1+A)\,,
\end{split}
\end{align}
which express the waiting times $A$ and $B$ as the average time to reach the
desired final state after a coin flip, plus the time for the coin flip
itself.  Thus in the equation for $A$, the first term accounts for the next
coin flip being H (which occurs with probability $\frac{1}{2}$) after which
the sequence HH has been generated.  The factor 2 counts the two coin flips
that are need to generate the sequence HH from scratch.  The second term
accounts for the next coin flip being T.  Again, the probability for this
event is $\frac{1}{2}$.  Once a T appears, the waiting time to generate an HH
sequence is $B$ by definition.  Consequently, the factor $(1+B)$ accounts for
the time spent in making a single coin flip plus the waiting time when the
sequence string starts with T.  Solving these two equations gives $A=5$ and
$B=7$.  Since H and T appear equiprobably in a long series of fair coin
flips, on average, the average waiting time for the sequence HH is
$T_{HH}=\frac{1}{2}(A+B)=6$.

For the waiting time $T_{HT}$, we now define
\vspace{-2mm}
\begin{itemize}
  \itemsep -0.5ex
\item $A$, the average waiting time for the sequence HT starting with H.
\item $B$, the average waiting time for the sequence HT starting with T.
\end{itemize}
Using the same reasoning as above, these two times obey the backward
equations
\begin{align}
  \begin{split}
    \label{AB2}
  A&= \tfrac{1}{2}\times 2 +\tfrac{1}{2}(1+A)\\
  B&= \tfrac{1}{2}(1+A) +\tfrac{1}{2}(1+B)\,,
\end{split}
\end{align}
\end{subequations}
from which $(A,B)=(3,5)$.  Again, because H and T appear equiprobably in a
long series of coin flips, $T_{HT}=\frac{1}{2}(A+B)=4$.

Why are these two times different?  The key lies in the second term on the
right in the first lines of Eqs.~\eqref{AB1} and \eqref{AB2}, which account
for a ``mistake''.  For example, in Eq.~\eqref{AB1}, if the next coin flip is
T, one has to ``start over'' to generate HH.  The soonest that the next HH
can happen immediately after a T is after two more coin flips.  In contrast,
in Eq.~\eqref{AB2}, if the next coin flip is H (again a mistake), the process
``starts over''.  Now, however, the next HT sequence can appear in after only
one more coin flip.

\subsection{Triplets}

We now generalize to triplet sequences.  The $2^3=8$ distinct triplets are
HHH, HHT, HTH, and THH and their counterparts obtained by
H$\leftrightarrow$T.  By left/right symmetry, the triplets HHT and THH have
identical statistics, so the only distinct sequences are HHH, HHT, and HTH.
Let $T_{\rm HHH}$ be the average waiting time to encounter the sequence with
three consecutive H's.  To compute this time, we define the auxiliary
restricted times: \vspace{-2mm}
\begin{itemize}
  \itemsep -0.5ex
\item $A$, the average waiting time for HHH when the current state is H;
\item $B$, the average waiting time for HHH when the current state is HH;
\item $C$, the average waiting time for HHH when the current state is T.
\end{itemize}
Following the same reasoning that led to Eqs.~\eqref{AB1}, the above times
satisfy 
\begin{align}
  \label{ABC1}
\begin{split}  
  A&=\tfrac{1}{2}(1+B)+\tfrac{1}{2}(1+C)\\
  B&=\tfrac{1}{2}\times 2+\tfrac{1}{2}(1+C)\\
  C&=\tfrac{1}{2}(1+A)+\tfrac{1}{2}(1+C)\,.
\end{split}
\end{align}
The first term in the equation for $B$ merits explanation.  From the state
HH, the desired sequence HHH is obtained with probability $\frac{1}{2}$, and
the time for this event is 2 because the time is measured starting
\emph{before} the second H has been added to the sequence.  The solution to
\eqref{ABC1} is $(A,B,C)=(13,9,15)$.  Since the probability to find an H or a
T are equal, the average waiting time to encounter the sequence HHH is just the
average of the times to find HHH when starting with an H or starting with a
T.  Thus $T_{\rm HHH}=\frac{1}{2}(A+C)=14$.

Similarly, let $T_{\rm HHT}$ be the average waiting time to encounter the sequence
HHT.  Now we introduce the auxiliary times: \vspace{-2mm}
\begin{itemize}
\itemsep -0.5ex
\item  $A$, the average waiting time for HHT when the current state
  is H;
\item $B$,  the average waiting time for HHT when the current state
  is HH;
\item $C$, the average waiting time for HHT when the current state
  is T.
\end{itemize}
These times satisfy 
\begin{align}
  \label{ABC2}
\begin{split}  
  A&=\tfrac{1}{2}(1+B)+\tfrac{1}{2}(1+C)\\
  B&=\tfrac{1}{2}\times 2+\tfrac{1}{2}(1+B)\\
  C&=\tfrac{1}{2}(1+A)+\tfrac{1}{2}(1+C)\,,
\end{split}
\end{align}
Again, there is a subtlety in the second equation: if the initial state is
HH, then after adding an H, the current state is still HH, so that that the
second term involves $B$.  This feature that the initial state consists of a
subsequence of length greater than one plays an increasing role for longer
sequences (see Appendices A and B).  The solution to \eqref{ABC2} is
$(A,B,C)=(7,3,9)$.  Thus the mean waiting time to encounter the sequence HHT
is the average of the times to find HHT after an H or after a T, which gives
$T_{\rm HHT}=\frac{1}{2}(A+C)=8$.

Finally, let $T_{\rm HTH}$ be the average waiting time to encounter the sequence HTH.
We introduce the auxiliary times:
\vspace{-2mm}
\begin{itemize}
  \itemsep -0.5ex
\item  $A$, the average waiting time for HTH when the current state
  is H;
\item $B$,  the average waiting time for HTH when the current state
  is HT;
\item $C$, the average waiting time for HHT when the current state
  is T.
\end{itemize}
These times satisfy 
\begin{align}
  \label{ABC3}
\begin{split}  
  A&=\tfrac{1}{2}(1+A)+\tfrac{1}{2}(1+B)\\
  B&=\tfrac{1}{2}\times 2+\tfrac{1}{2}(1+C)\\
  C&=\tfrac{1}{2}(1+A)+\tfrac{1}{2}(1+C)\,,
\end{split}
\end{align}
with solutions $(A,B,C)=(9,7,11)$.  The mean waiting time to encounter the
sequence HTH is the average of the times to find HTH after an H or after a T,
which gives $T_{\rm HTH}=\frac{1}{2}(A+C)=10$.  To summarize,
$T_{\rm HHH}=14$, $T_{\rm HHT}=8$, and $T_{\rm HTH}=10$, in agreement with
known results.  The corresponding results for quartet and quintet sequences
are given in~\ref{app:4} and \ref{app:5}.

\section{Mean-Square Waiting Times}
\label{sec:tsq}

We now extend the backward Kolmogorov approach to compute the mean-square
waiting time for specific sequences.  As we will show in the next section,
deriving the moment generating function requires less computation than the
second moment, and it becomes increasingly laborious to directly derive
moments beyond the second.  Thus we present the calculation of the second
moment for doublet sequences as an illustration only, and then proceed to the
moment generating functions for doublets and longer sequences.

Let us start with the sequence HH.  In analogy with the discussion of the
mean waiting time, we now introduce the auxiliary restricted mean-square
times: \vspace{-2mm}
\begin{itemize}
  \itemsep -0.5ex
\item $A_2$, the mean-square waiting time for HH starting with an H.
\item $B_2$, the mean-square waiting time for HH starting with a T.
\end{itemize}
These two quantities obey the backward equations
\begin{subequations}
\begin{align}
\begin{split}
  \label{AB1sq}
  A_2&= \tfrac{1}{2}\times 4 +\tfrac{1}{2}\langle 1+\mathcal{B}\rangle^2\\
  B_2&= \tfrac{1}{2}\langle 1+\mathcal{B}\rangle^2 +\tfrac{1}{2} \langle 1+\mathcal{A}\rangle^2\,.
\end{split}
\end{align}
Here the notations $\mathcal{A}$ and $\mathcal{B}$ denote the time to reach
the sequence HH from a specific sequence realization when starting from an H
or a T respectively, and the angle brackets denote an average over all
sequences.  According to this notation, $A=\langle \mathcal{A}\rangle$ and
$A_2=\langle \mathcal{A}^2\rangle$.

We expand the quadratic inside the angle brackets to give
\begin{align}
\begin{split}
  \label{AB1sq2}
  A_2&= 2 +\tfrac{1}{2}( 1+2B +B_2)\\
  B_2&= \tfrac{1}{2}(1+2B+B_2) +\tfrac{1}{2} (1+2A+A_2)\,,
\end{split}
\end{align}
\end{subequations}
and then substitute in the solutions $A=5$ and $B=7$ from Eq.~\eqref{AB1} to
obtain $A_2=45$ and $B_2=71$.  Since H and T occur with equal probability,
the mean-square waiting time to find the sequence HH is the average of $A_2$
and $B_2$: $T^2_{\rm HH} = 58$.

For the sequence HT, we introduce
\vspace{-2mm}
\begin{itemize}
  \itemsep -0.5ex
\item $A_2$, the mean-square waiting time for HT starting with H.
\item $B_2$, the mean-square waiting time for HT starting with T.
\end{itemize}
These two times obey the backward equations
\begin{align}
  \begin{split}
    \label{AB2sq}
  A_2&= \tfrac{1}{2}\times 4 +\tfrac{1}{2}\langle 1+A\rangle^2= 2 +\tfrac{1}{2}(1+2A +A_2) \\
  B_2&= \tfrac{1}{2}\langle 1+A\rangle^2 +\tfrac{1}{2}\langle 1+B\rangle^2= \tfrac{1}{2}(1+2A+A_2) +\tfrac{1}{2} (1+2B+B_2)\,.
\end{split}
\end{align}
Using $A=3$ and $B=5$ from Eq.~\eqref{AB2}, we now find $A_2=11$ and
$B_2=29$.  Then the mean-square waiting time for the sequence HT is
$T^2_{\rm HT}=\frac{1}{2}(A_2+B_2) = 20$.

Clearly, this same approach can be extended to higher moments, but the
calculations become progressively more tedious because the equations for the
$n^{\rm th}$ moment involves all lower-order moments.  As shown below,
however, we can compute the moment generating function with less
computational labor than that for the second moment, and from this generating
function all moments are obtained by a simple Taylor series expansion.

\section{The Moment Generating Function}
\label{sec:mgf}

\subsection{Doublet Sequence HH}

To begin, we define the moment generating functions
\begin{align*}
  G_A = \langle e^{-s\mathcal{A}}\rangle\qquad\qquad  G_B = \langle e^{-s\mathcal{B}}\rangle\,,
\end{align*}
where again $\mathcal{A}$ and $\mathcal{B}$ denote the time to reach the
sequence HH from a specific sequence realization when starting from an H or a
T respectively.  These generating functions are especially useful because
they contain all moments of the waiting time by a Taylor series expansion.
For example, the moments of the time $A$ to encounter the sequence HH when
starting with an H are
\begin{align*}
  G_A = 1 - s\,A +\frac{1}{2}\, s^2 A_2 - \frac{1}{3!}\, s^2 A_3 +\ldots
\end{align*}

For the sequence HH, the backward Kolmogorov equation for the moment
generating functions are obtained by taking each of the terms in
Eqs.~\eqref{AB1} and putting it inside the exponential like so:
$e^{-s \times \text{each\ term}}$.  This immediately gives
\begin{align}
\begin{split}
\label{GHH}
G_A &=\tfrac{1}{2}\langle e^{-2s}\rangle +\tfrac{1}{2}\langle e^{-s(1+\mathcal{B})}\rangle
= \tfrac{1}{2}(e^{-2s}+ G_B\,e^{-s})\\
G_B &=\tfrac{1}{2}\langle e^{-s(1+\mathcal{B})}\rangle +\tfrac{1}{2}\langle
e^{-s(1+\mathcal{A})}\rangle = \tfrac{1}{2} (G_B\,e^{-s}+ G_A\,e^{-s})\,.
\end{split}
\end{align}
The structure of these equations mirror those of Eqs.~\eqref{AB1} for the
mean waiting time.  Because of this close correspondence, solving the
backward equations for the moment generating functions has the same degree of
difficulty as solving for the mean waiting times.  The solution to
\eqref{GHH} is
\begin{align}
\label{GAGB}
  G_A= \frac{(2\,e^s-1)e^{-s}}{4\,e^{2s}-2\,e^s-1}\qquad \qquad 
  G_B= \frac{e^{-s}}{4\,e^{2s}-2\,e^s-1}\,.
\end{align}
Taylor expanding these two generating functions gives
\begin{align}
  G_A = 1- 5\, s +\frac{45}{2}s^2 - \frac{629}{3!} s^3 +\ldots\qquad
  G_B = 1- 7\, s +\frac{71}{2}s^2 - \frac{1015}{3!} s^3 +\ldots\,,
\end{align}
from which $T_{HH} = \frac{1}{2}(A+B) = 6$,
$T^2_{HH} = 2\times \frac{1}{2}(A_2+B_2) = 58$, and
$T^3_{HH} = 3!\times \frac{1}{2}(A_3+B_3) = 822$, etc.

Finally, we note that the moment generating functions in \eqref{GAGB} both
have simple poles at $s^*\approx -0.21194$. This implies that the long-time
tail of the distribution of waiting times, $P(t)$, has an exponential decay
of the form $e^{-|s^*|t} = e^{-t/\tau}$, with $\tau\approx 4.718$.  Even though
the times $T_{HH}$ and $T_{HT}$ are numerically different, both times are
governed by a single characteristic scale.

\subsection{Doublet Sequence HT}

Building on the above perspective, we write the backward equations for the
moment generating functions for the sequence HT by merely reading off from
Eqs.~\eqref{AB2}:
\begin{align}
\begin{split}
\label{GHT}
G_A &= \tfrac{1}{2}(e^{-2s}+ G_A\,e^{-s})\\
G_B & = \tfrac{1}{2} (G_B\,e^{-s}+ G_A\,e^{-s})\,,
\end{split}
\end{align}
with solutions
\begin{align}
  G_A= \frac{e^{-s}}{2\,e^s-1}\qquad \qquad 
  G_B= \frac{e^{-s}}{(2\,e^s-1)^2}\,.
\end{align}
Taylor expanding these two generating functions gives
\begin{align}
  G_A = 1- 3\, s +\frac{11}{2}s^2 - \frac{51}{3!} s^3 +\ldots\qquad
  G_B = 1- 5\, s +\frac{29}{2}s^2 - \frac{197}{3!} s^3 +\ldots\,,
\end{align}
from which $T_{HT} = \frac{1}{2}(A+B) = 4$,
$T^2_{HT} = 2\times \frac{1}{2}(A_2+B_2) = 20$, and
$T^3_{HT} = 6\times \frac{1}{2}(A_3+B_3) = 124$, etc.

Corresponding results for triplet sequences are given in \ref{app:3}.

\section{Simple Arbitrary Length Sequences}

\subsection{Mean waiting time for $n$ consecutive H's}
\label{subsec:n}

While the calculational details for longer sequences are straightforward,
they become progressively more tedious as the sequence length is increased.
However, for the sequence of $n$ consecutive H's, the equations for the
restricted times are sufficiently systematic in character that they can be
readily solved.  To this end, we first define the following set of restricted
times: \vspace{-2mm}
\begin{itemize}
  \itemsep -0.5ex
\item $A_k$, the average waiting time for $n$H starting from $k$ consecutive
  H's;
\item $B$, the average waiting time for $n$H starting from T.
\end{itemize}
These times satisfy 
\begin{align}
\label{nA}
\begin{split}
  A_1 &= \tfrac{1}{2}(1+A_2)+\tfrac{1}{2}(1+B)\\
  A_2 &= \tfrac{1}{2}(1+A_3)+\tfrac{1}{2}(1+B)\\
      & \vdots\\
  A_{n-2}&=  \tfrac{1}{2}(1+A_{n-1})+\tfrac{1}{2}(1+B)\\
  A_{n-1}&=  \tfrac{1}{2}\times 2+   \tfrac{1}{2}(1+B)\\
  B&= \tfrac{1}{2}(1+A_1)+\tfrac{1}{2}(1+B)\,.
\end{split}
\end{align}
% From the last equation, we have $B=2+A_1$, while from the penultimate
% equation we can replace the factor $\frac{1}{2}(1+B)$ everywhere with
% $A_{n-1}-1$.  Thus the first $n-2$ of Eqs.~\eqref{nA} become
% \begin{align}
% \label{nAa}
% \begin{split}
%   A_1 &= \tfrac{1}{2}(1+A_2)+A_{n-1}-1\\
%   A_2 &= \tfrac{1}{2}(1+A_3)+A_{n-1}-1\\
%       & \vdots\\
%   A_{n-3}&=  \tfrac{1}{2}(1+A_{n-2})+A_{n-1}-1\\
%   A_{n-2}&=  \tfrac{1}{2}(1+A_{n-1})+A_{n-1}-1\\
% \end{split}
% \end{align}
% \end{subequations}

% Notice that the equation for $A_{n-2}$ is expressed in terms of $A_{n-1}$
% only.  We then substitute this expression for $A_{n-2}$ in the equation for
% $A_{n-3}$, so that $A_{n-3}$ can also be expressed in terms of $A_{n-1}$
% only.  
%Continuing this procedure, we find
% \begin{subequations}
% \begin{align}
%   A_{n-k} = -\frac{2^{k-1}-1}{2^{k-1}}+\frac{2^k-1}{2^{k-1}} A_{n-1}\,.
% \end{align}
% In particular
% \begin{align}
%   \label{A1}
%   A_1= -\frac{2^{n-2}-1}{2^{n-2}}+\frac{2^{n-1}-1}{2^{n-2}} A_{n-1}\,.
% \end{align}
% \end{subequations}
% From the original equation for $A_{n-1}$, we eliminate $B$ in favor of $A_1$
% and obtain $A_{n-1}=\frac{5}{2}+\frac{1}{2}A_1$. 
Because of the linear and recursive structure of Eqs.~\eqref{nA} they can be
solved one by one, and the final result, for the waiting time to
$n$H when starting from an H or a T respectively, is
\begin{align}
  A_1= 2^{n+2}-3 \qquad\qquad B=2^{n+2}-1\,.
\end{align}
Finally, $T_{\rm nH}$ is the average of these two waiting times:
\begin{align}
  \label{T-nH}
  T_{\rm nH}=\tfrac{1}{2}(A_1+B)=2^{n+2}-2\,.
\end{align}

\subsection{$n$ Consecutive (HT)'s}

We can carry out a similar calculation for the sequence of $n$ consecutive
(HT)'s.  Here, we first define the following set of restricted times:
\vspace{-2mm}
\begin{itemize}
  \itemsep -0.5ex
\item $A_{2k-1}$, the average waiting time for $n$(HT) starting from
  $(k-1)$(HT)H;
\item $A_{2k}$, the average waiting time for $n$(HT) starting from $k$(HT);
\item $B$, the average waiting time for $n$(HT) starting from T.
\end{itemize}

These times satisfy 
\begin{align}
\label{nHT}
\begin{split}
  A_1 &= \tfrac{1}{2}(1+A_2)+\tfrac{1}{2}(1+A_1)\\
  A_2 &= \tfrac{1}{2}(1+A_3)+\tfrac{1}{2}(1+B)\\
  A_3 &= \tfrac{1}{2}(1+A_4)+\tfrac{1}{2}(1+A_1)\\
  A_4 &= \tfrac{1}{2}(1+A_5)+\tfrac{1}{2}(1+B)\\
      & \vdots\\
  A_{2n-3}&=  \tfrac{1}{2}(1+A_{2n-2})+\tfrac{1}{2}(1+A_1)\\
  A_{2n-2}&=  \tfrac{1}{2}(1+A_{2n-1})+\tfrac{1}{2}(1+B)\\
  A_{2n-1}&=  \tfrac{1}{2}\times 2+\tfrac{1}{2}(1+A_1)\\
    B&= \tfrac{1}{2}(1+A_1)+\tfrac{1}{2}(1+B)\,.
\end{split}
\end{align}
Solving these equations recursively, we find
% The last two equations give
% \begin{align}
% \label{A1-B}  \tfrac{1}{2}(1+A_1)= A_{2n-1}\qquad\qquad \tfrac{1}{2}(1+B) =A_{2n-1}\,.
% \end{align}
% We now eliminate $A_1$ and $B$ from the rest of Eqs.~\eqref{nHT}.  After
% these eliminations, we can then recursively solve for $A_{2n-2}$,
% $A_{2n-3},\ldots$ in terms of $A_{2n-1}$ and obtain
% \begin{align}
% \label{A1-alt}
%   A_{2n-k}= -\frac{S_{k}}{2^{k-1}}+ \frac{2^k-1}{2^{k-1}}\,A_{2n-1}\,.
% \end{align}
% where
% \begin{align*}
%   S_k=\sum_{j=0}^{(k-3)/2}4^j \,.
% \end{align*}
% We now use $k=2n-1$ in \eqref{A1-alt} to obtain $A_1$ in terms of
% $A_{2n-1}$, and then use the first of \eqref{A1-B} to eliminate $A_{2n-1}$ in
% favor of $A_1$ and ultimately solve for $A_1$.  After straightforward
% algebra, the final result is
\begin{align}
  A_1 = \tfrac{4}{3}(2^{2n}-1)-1\qquad\qquad B= 2+A_1\,.
\end{align}
The average waiting time $T_{n\rm(HT)}=\frac{1}{2}(A_1+B)=A_1+1$ now is
\begin{align}
\label{T-nHT}    
  T_{n\rm (HT)}= \tfrac{4}{3}(2^{2n}-1)\,.
\end{align}

It is instructive to compare the times $T_{n\rm H}$ and $T_{n\rm(HT)}$.  The
fair comparison is between $T_{2n\rm H}$ and $T_{n\rm(HT)}$; i.e., between
strings of the same length.  Asymptotically, Eq.~\eqref{T-nH} gives
$T_{2n\rm H}\sim 4\cdot 2^{2n}$, while \eqref{T-nHT} gives
$T_{n\rm(HT)}\sim \frac{4}{3}\cdot 2^{2n}$.  One has to wait three times as
long, on average, to encounter a sequence of $2n$ H's in a row compared to a
sequence of $n$ (HT)'s in a row.

\subsection{Moment generating function for $n$ consecutive H's}
\label{subsec:mgf-n}

We now use the approach outlined in Sec.~\ref{sec:mgf} to compute the moment
generating function for the occurrence of $n$ consecutive H's. Following the
notation of Sec.~\ref{subsec:n}, we define $G_k$ as the moment generating
function for the time to reach the state of nH's when the sequence starts
with $k$ consecutive H's, while $G_B$ is the moment generating function to
reach the state nH when the sequence starts with a T.  In close analogy with
Eqs.~\eqref{nA}, these moment generating functions satisfy
\begin{align}
\label{nA-mgf}
\begin{split}
  G_1 &= x\, G_2+x \,G_B\\
  G_2 &= x\,G_3+x\,G_B\\
      & \vdots\\
  G_{n-2}&=  x\,G_{n-1}+x\,G_B\\
  G_{n-1}&=  y+x\,G_B\\
  G_B&= x\,G_1+x\,G_B\,,
\end{split}
\end{align}
where for notational simplicity we introduce $x\equiv\tfrac{1}{2}\,e^{-s}$
and $y\equiv \tfrac{1}{2} \,e^{-2s}$.  The last equation gives
$G_B=x\,G_1/(1-x)$, while from the penultimate equation we can replace the
factor $x\,G_B$ everywhere with $G_{n-1}-y$.  

Following similar steps as those used to solve \eqref{nA}, the moment
generating functions $G_1$ and $G_B$ are
\begin{align}
  G_1 = \frac{x^{n-1}(1-x)^2 y}{(1-x)^2-x^2(1-x^n)}\qquad
   G_B = \frac{x^{n}(1-x) y}{(1-x)^2-x^2(1-x^n)}\,.
\end{align}
The moment generating function for the time $T_{n\rm H}$ to encounter the
sequence $n$H is the average of $G_1$ and $G_B$; that is
\begin{align}
  \tfrac{1}{2}(G_1+G_B)=\frac{1}{2} \frac{x^{n-1}(1-x) y}{(1-x)^2-x^2(1-x^n)}
\end{align}
Finally, we may expand this generating function in a power series to obtain
the moments of the time to reach the sequence $n$H. The first few moments are:
\begin{align}
  \begin{split}
\langle T_{n\rm H}\rangle &= 1!\cdot 2^{n+2}-2\\
  \langle T_{n\rm H}^2\rangle &= 2!\cdot 2^{2n+4}- n\cdot 2^{n+2}-7\cdot 2^{n+1}+1\\
  \langle T_{n\rm H}^3\rangle &\simeq 3!\cdot 2^{3n+6}\\
   \langle T_{n\rm H}^4\rangle &\simeq 4!\cdot 2^{4n+8}\\
\end{split}
\end{align}

\section{Concluding Comments}

While many of the results given here are already quite well known, the
backward Kolmogorov approach provides a fresh perspective to calculate
average waiting times for specific sequences of H's and T's in a long string
of repeated flips of a fair coin.  Once one understands the underlying idea
of the Kolmogorov approach, computing waiting times for specific sequences is
straightforward and direct.

Another important aspect of this approach is that it also allows one to
compute any \emph{functional} of the waiting time, such as higher moments,
and even the characteristic function, $\langle \exp(-sT)\rangle$.  We showed
how to compute the moment generating function for short specific sequences of
heads and tails, from which arbitrary moments of the waiting time for these
sequences can easily be derived.  This approach can be readily extended to
longer sequences.  These results about higher moments appear to have not been
treated previously.  An open challenge is whether there exists a simple
approach of the spirit given in~\cite{A11}, that allows one to compute the
moment generating function for any sequence of arbitrary length.

The surprising outcome of repeated fair coin flips is that the average
waiting times for specific sequences of H's and T's of the same length are
different even though the average frequency of these two sequences are the
same.  The effect is especially pronounced for a long string of $2n$ H's
compared to the string of $n$ (HT)'s.  For large $n$, one has to wait three
times longer to encounter the former sequence compared to the latter.

As a final note, although our approach unambiguously demonstrates the
existence of distinct waiting times for distinct fixed-length
sequences, this seemingly paradoxical phenomenon requires careful
thought to appreciate intuitively.

I thank David Atkinson and Porter Johnson for helpful suggestions while this
manuscript was being written, Paul Krapivsky and Michael Mauboussin for their
encouragement, and Ivan Corwin for helpful advice.  I also gratefully
acknowledge financial support from NSF Grants DMR-1608211 and DMR-1910736.

\appendix

\section{Calculational Details for Quartet Sequences}
\label{app:4}

The six distinct quartets are HHHH, HHHT, HHTH, HHTT, HTHT, and HTTH.  The
calculation $T_{\rm HHHH}$ was given in Sec.~\ref{subsec:n} and here we continue
with $T_{\rm HHHT}$.  For $T_{\rm HHHT}$, we define $A$, $B$, $C$, and $D$ as
the average waiting time for HHHT when the current state is H, HH, HHH, and T,
respectively.  These times satisfy
\begin{align}
  \label{ABCD2}
\begin{split}  
  A&=\tfrac{1}{2}(1+B)+\tfrac{1}{2}(1+D)\\
  B&=\tfrac{1}{2}(1+C)+\tfrac{1}{2}(1+D)\\
  C&=\tfrac{1}{2}\times 2+\tfrac{1}{2}(1+C)\\
  D&=\tfrac{1}{2}(1+A)+\tfrac{1}{2}(1+D)\,,
\end{split}
\end{align}
with solution $(A,B,C,D)=(15,11,3,17)$, from which
$T_{\rm HHHT}=\frac{1}{2}(A+D)=16$.

For $T_{\rm HHTH}$, we define $A$, $B$, $C$, and $D$ as the average waiting
time for HHTH when the current state is H, HH, HHT, and T, respectively.
These times satisfy
\begin{align}
  \label{ABCD3}
\begin{split}  
  A&=\tfrac{1}{2}(1+B)+\tfrac{1}{2}(1+D)\\
  B&=\tfrac{1}{2}(1+B)+\tfrac{1}{2}(1+C)\\
  C&=\tfrac{1}{2}\times 2+\tfrac{1}{2}(1+D)\\
  D&=\tfrac{1}{2}(1+A)+\tfrac{1}{2}(1+D)\,,
\end{split}
\end{align}
with solution $(A,B,C,D)=(17,13,11,19)$, from which
$T_{\rm HHTH}=\frac{1}{2}(A\!+\!D)=18$.

For $T_{\rm HHTT}$, we define $A$, $B$, $C$, $D$ as the average waiting time
for HHTT when the current state is H, HH, HHT, and T, respectively.  These
times satisfy
\begin{align}
  \label{ABCD4}
\begin{split}  
  A&=\tfrac{1}{2}(1+B)+\tfrac{1}{2}(1+D)\\
  B&=\tfrac{1}{2}(1+B)+\tfrac{1}{2}(1+C)\\
  C&=\tfrac{1}{2}\times 2+\tfrac{1}{2}(1+A)\\
  D&=\tfrac{1}{2}(1+A)+\tfrac{1}{2}(1+D)\,,
\end{split}
\end{align}
with solution $(A,B,C,D)=(15,11,9,17)$, from which
$T_{\rm HHTT}=\frac{1}{2}(A+D)=16$.

For $T_{\rm HTHT}$, we define $A$, $B$, $C$, $D$ as the average waiting time
for HTHT when the current state is H, HT, HTH, and T, respectively.  These
times satisfy
\begin{align}
  \label{ABCD5}
\begin{split}  
  A&=\tfrac{1}{2}(1+A)+\tfrac{1}{2}(1+B)\\
  B&=\tfrac{1}{2}(1+C)+\tfrac{1}{2}(1+D)\\
  C&=\tfrac{1}{2}\times 2+\tfrac{1}{2}(1+A)\\
  D&=\tfrac{1}{2}(1+A)+\tfrac{1}{2}(1+D)\,,
\end{split}
\end{align}
with solution $(A,B,C,D)=(19,17,11,21)$, from which
$T_{\rm HTHT}=\frac{1}{2}(A+D)=20$.

Finally, for $T_{\rm HTTH}$, we define $A$, $B$, $C$, $D$ as the
average waiting time for HTTH when the current state is H, HT, HTT, and T,
respectively.  These times satisfy
\begin{align}
  \label{ABCD6}
\begin{split}  
  A&=\tfrac{1}{2}(1+A)+\tfrac{1}{2}(1+B)\\
  B&=\tfrac{1}{2}(1+A)+\tfrac{1}{2}(1+C)\\
  C&=\tfrac{1}{2}\times 2+\tfrac{1}{2}(1+D)\\
  D&=\tfrac{1}{2}(1+A)+\tfrac{1}{2}(1+D)\,,
\end{split}
\end{align}
with solution $(A,B,C,D)=(17,15,11,19)$, from which
$T_{\rm HTTH}=\frac{1}{2}(A+D)=18$.

In summary, the quartet average waiting times in reverse time order are
$T_{\rm 4H}=30$, $T_{\rm HTHT}=20$, $T_{\rm HHTH}=T_{\rm HTTH}=18$,
$T_{\rm HHHT}= T_{\rm HHTT}=16$.

\section{Quintet Sequences}
\label{app:5}

The nine distinct quintets are: HHHHH, HHHHT, HHHTH, HHTHH, HHHTT, HHTHT,
HTHHT, HTHTH, and HTTHH.  There are additional non-independent sequences that
are obtained by either the interchange $H\leftrightarrow T$ or by reading the
above sequences in reverse order.  Again, the calculation $T_{\rm HHHHH}$ was
given in Sec.~\ref{subsec:n} and we continue with $T_{\rm HHHHT}$.  For
$T_{\rm HHHHT}$, we define $A$, $B$, $C$, $D$, $E$ as the average waiting time for
HHHHT when the current state is H, HH, HHH, HHHH, and T, respectively.  These
times satisfy
\begin{align}
  \label{ABCDE1}
\begin{split}  
  A&=\tfrac{1}{2}(1+B)+\tfrac{1}{2}(1+E)\\
  B&=\tfrac{1}{2}(1+C)+\tfrac{1}{2}(1+E)\\
  C&=\tfrac{1}{2}(1+D)+\tfrac{1}{2}(1+E)\\
  D&=\tfrac{1}{2}\times 2+\tfrac{1}{2}(1+D)\\
  E&=\tfrac{1}{2}(1+A)+\tfrac{1}{2}(1+E)\\
\end{split}
\end{align}
with solution $(A,B,C,D,E)=(31,27,19,3,33)$, and we obtain
$T_{\rm HHHHT}=\frac{1}{2}(A+E)=32$.

For $T_{\rm HHHTH}$, we define $A$, $B$,
$C$, $D$, $E$ as the average waiting time for HHHTH when the current state is H, HH,
HHH, HHHT, and T, respectively.  These times satisfy
\begin{align}
  \label{ABCDE2}
\begin{split}  
  A&=\tfrac{1}{2}(1+B)+\tfrac{1}{2}(1+E)\\
  B&=\tfrac{1}{2}(1+C)+\tfrac{1}{2}(1+E)\\
  C&=\tfrac{1}{2}(1+C)+\tfrac{1}{2}(1+D)\\
  D&=\tfrac{1}{2}\times 2+\tfrac{1}{2}(1+E)\\
  E&=\tfrac{1}{2}(1+A)+\tfrac{1}{2}(1+E)\\
\end{split}
\end{align}
with solution $(A,B,C,D,E)=(33,29,21,19,35)$, and we obtain
$T_{\rm HHHTH}=\frac{1}{2}(A+E)=34$.

For $T_{\rm HHTHH}$, we define $A$, $B$, $C$, $D$, $E$ as the average waiting time
for HHTHH when the current state is H, HH, HHT, HHTH, and T, respectively.
These times satisfy
\begin{align}
  \label{ABCDE3}
\begin{split}  
  A&=\tfrac{1}{2}(1+B)+\tfrac{1}{2}(1+E)\\
  B&=\tfrac{1}{2}(1+B)+\tfrac{1}{2}(1+C)\\
  C&=\tfrac{1}{2}(1+D)+\tfrac{1}{2}(1+E)\\
  D&=\tfrac{1}{2}\times 2+\tfrac{1}{2}(1+E)\\
  E&=\tfrac{1}{2}(1+A)+\tfrac{1}{2}(1+E)\\
\end{split}
\end{align}
with solution $(A,B,C,D,E)=(37,33,31,21,39)$, and we obtain
$T_{\rm HHTHH}=\frac{1}{2}(A+E)=38$.

For $T_{\rm HHHTT}$, we define $A$, $B$, $C$, $D$, $E$ as the average waiting time
for HHHTT when the current state is H, HH, HHH, HHHT, and T, respectively.
These times satisfy
\begin{align}
  \label{ABCDE4}
\begin{split}  
  A&=\tfrac{1}{2}(1+A)+\tfrac{1}{2}(1+E)\\
  B&=\tfrac{1}{2}(1+C)+\tfrac{1}{2}(1+E)\\
  C&=\tfrac{1}{2}(1+C)+\tfrac{1}{2}(1+D)\\
  D&=\tfrac{1}{2}\times 2+\tfrac{1}{2}(1+A)\\
  E&=\tfrac{1}{2}(1+A)+\tfrac{1}{2}(1+E)\\
\end{split}
\end{align}
with solution $(A,B,C,D,E)=(31,27,19,17,33)$, and we obtain
$T_{\rm HHHTT}=\frac{1}{2}(A+E)=32$.

For $T_{\rm HHTHT}$, we define $A$, $B$,
$C$, $D$, $E$ as the average waiting time for HHTHT when the current state is H, HH,
HHT, HHTH, and T, respectively.  These times satisfy
\begin{align}
  \label{ABCDE5}
\begin{split}  
  A&=\tfrac{1}{2}(1+B)+\tfrac{1}{2}(1+E)\\
  B&=\tfrac{1}{2}(1+B)+\tfrac{1}{2}(1+C)\\
  C&=\tfrac{1}{2}(1+D)+\tfrac{1}{2}(1+E)\\
  D&=\tfrac{1}{2}\times 2+\tfrac{1}{2}(1+B)\\
  E&=\tfrac{1}{2}(1+A)+\tfrac{1}{2}(1+E)\\
\end{split}
\end{align}
with solutions $(A,B,C,D,E)=(31,27,25,15,33)$, and we obtain
$T_{\rm HHTHT}=\frac{1}{2}(A+E)=32$.

For $T_{\rm HTHHT}$, we define $A$, $B$, $C$, $D$, $E$ as the average waiting time
for HTHHT when the current state is H, HT, HTH, HTHH, and T, respectively.
These times satisfy
\begin{align}
  \label{ABCDE6}
\begin{split}  
  A&=\tfrac{1}{2}(1+A)+\tfrac{1}{2}(1+B)\\
  B&=\tfrac{1}{2}(1+C)+\tfrac{1}{2}(1+E)\\
  C&=\tfrac{1}{2}(1+D)+\tfrac{1}{2}(1+B)\\
  D&=\tfrac{1}{2}\times 2+\tfrac{1}{2}(1+A)\\
  E&=\tfrac{1}{2}(1+A)+\tfrac{1}{2}(1+E)\\
\end{split}
\end{align}
with solution $(A,B,C,D,E)=(35,33,27,19,37)$, and we obtain
$T_{\rm HTHHT}=\frac{1}{2}(A+E)=36$.

For $T_{\rm HTHTH}$, we define $A$, $B$, $C$, $D$, $E$ as the average waiting time
for HTHTH when the current state is H, HT, HTH, HTHT, and T, respectively.
These times satisfy
\begin{align}
  \label{ABCDE7}
\begin{split}  
  A&=\tfrac{1}{2}(1+A)+\tfrac{1}{2}(1+B)\\
  B&=\tfrac{1}{2}(1+C)+\tfrac{1}{2}(1+E)\\
  C&=\tfrac{1}{2}(1+A)+\tfrac{1}{2}(1+D)\\
  D&=\tfrac{1}{2}\times 2+\tfrac{1}{2}(1+E)\\
  E&=\tfrac{1}{2}(1+A)+\tfrac{1}{2}(1+E)\\
\end{split}
\end{align}
with solution $(A,B,C,D,E)=(41,39,33,23,43)$, and we obtain
$T_{\rm HTHTH}=\frac{1}{2}(A+E)=42$.

For $T_{\rm HTTHH}$, we define $A$, $B$,
$C$, $D$, $E$ as the average waiting time for HTTHH when the current state is H, HT,
HTT, HTTH, and T, respectively.  These times satisfy
\begin{align}
  \label{ABCDE8}
\begin{split}  
  A&=\tfrac{1}{2}(1+A)+\tfrac{1}{2}(1+B)\\
  B&=\tfrac{1}{2}(1+A)+\tfrac{1}{2}(1+C)\\
  C&=\tfrac{1}{2}(1+D)+\tfrac{1}{2}(1+E)\\
  D&=\tfrac{1}{2}\times 2+\tfrac{1}{2}(1+B)\\
  E&=\tfrac{1}{2}(1+A)+\tfrac{1}{2}(1+E)\\
\end{split}
\end{align}
with solution $(A,B,C,D,E)=(33,31,27,17,35)$, and we obtain
$T_{\rm HTTHH}=\frac{1}{2}(A+E)=34$.

In summary, the quintet average waiting times in reverse time order are:
$T_{\rm 5H}=62$, $T_{\rm HTHTH}=42$, $T_{\rm HHTHH}=38$, $T_{\rm HTHHT}=36$,
$T_{\rm HHHTH}=T_{\rm HTTHH}=34$, $T_{\rm HHHTH}=T_{\rm HHTHT}=32$.  All the
results for quartets and quintets agree with those given in \cite{JA10}.

\section{Moment Generating Functions for Triplet Sequences}
\label{app:3}

For the sequence HHH, we define the moment generating functions
\begin{align*}
  G_A = \langle s^{-s\mathcal{A}}\rangle\qquad\qquad  G_B = \langle
  e^{-s\mathcal{B}}\rangle \qquad\qquad  G_C = \langle e^{-s\mathcal{C}}\rangle\,,
\end{align*}
where $\mathcal{A}$, $\mathcal{B}$, and $\mathcal{C}$ denote the time to
reach the sequence HHH from a specific sequence realization when starting
from H, HH, or T respectively.  Reading off from Eqs.~\eqref{ABC1}, the
backward equations for these moment generating functions are:
\begin{align}
\begin{split}
\label{GHHH}
G_A &= \tfrac{1}{2}(G_B\,e^{-s}+ G_C\,e^{-s})\\
G_B & = \tfrac{1}{2}(e^{-2s}+ G_C\,e^{-s})\\
G_C & = \tfrac{1}{2}(G_A\,e^{-s}+ G_C\,e^{-s})\,,
\end{split}
\end{align}
with solutions
\begin{align}
\begin{split}
  G_A&= \frac{e^{-s}(2\,e^s-1)}{8\,e^{3s}-4\,e^{2s}-2\,e^{s}-1}\\
  G_B&= \frac{e^{-s}(4\,e^{2s}-2\,e^{2}-1)}{8\,e^{3s}-4\,e^{2s}-2\,e^{s}-1}\\
  G_C&= \frac{e^{-s}}{8\,e^{3s}-4\,e^{2s}-2\,e^{s}-1}\,.
\end{split}
\end{align}
Taylor expanding these generating functions gives
\begin{align*}
  G_A &= 1- 13 s +\frac{309}{2}s^2 - \frac{11053}{3!} s^3 +\ldots\\
  G_B &= 1- 9 s +\frac{201}{2}s^2 - \frac{7161}{3!} s^3 +\ldots\\
  G_C &= 1- 15 s +\frac{367}{2}s^2 - \frac{13167}{3!} s^3 +\ldots\\
\end{align*}
from which $T_{HHH} = \frac{1}{2}(A+C) = 14$,
$T^2_{HHH} = 2\times \frac{1}{2}(A_2+C_2) = 338$, and
$T^3_{HHH} = 6\times \frac{1}{2}(A_3+C_3) = 12110$, etc.

For the sequence HHT, we  read off from Eqs.~\eqref{ABC2} to give the
backward equations for the moment generating functions:
\begin{align}
\begin{split}
\label{GHHT}
G_A &= \tfrac{1}{2}(G_B\,e^{-s}+ G_C\,e^{-s})\\
G_B & = \tfrac{1}{2}(e^{-2s}+ G_B\,e^{-s})\\
G_C & = \tfrac{1}{2}(G_A\,e^{-s}+ G_C\,e^{-s})\,,
\end{split}
\end{align}
with solutions
\begin{align}
\begin{split}
  G_A&= \frac{e^{-s}}{4\,e^{2s}-2\,e^{s}-1}= 1- 7 s +\frac{71}{2}s^2 - \frac{1015}{3!} s^3 +\ldots\\
  G_B&= \frac{e^{-s}}{2\,e^{s}-1} =1- 3 s +\frac{11}{2}s^2 - \frac{51}{3!} s^3 +\ldots\\
  G_C&= \frac{e^{-s}}{(4\,e^{2s}-2\,e^{s}-1)(2\,e^2-1)} =1- 9 s +\frac{105}{2}s^2 - \frac{1593}{3!} s^3 +\ldots\,,
\end{split}
\end{align}
from which $T_{HHT} = \frac{1}{2}(A+C) = 8$,
$T^2_{HHT} = 2\times \frac{1}{2}(A_2+C_2) = 88$, and
$T^3_{HHT} = 6\times \frac{1}{2}(A_3+C_3) = 1304$, etc.

For the sequence HTH, we  read off from Eqs.~\eqref{ABC3} to give the
backward equations for the moment generating functions:
\begin{align}
\begin{split}
\label{GHTH}
G_A &= \tfrac{1}{2}(G_A\,e^{-s}+ G_B\,e^{-s})\\
G_B & = \tfrac{1}{2}(e^{-2s}+ G_C\,e^{-s})\\
G_C & = \tfrac{1}{2}(G_A\,e^{-s}+ G_C\,e^{-s})\,,
\end{split}
\end{align}
with solutions
\begin{align}
\begin{split}
  G_A&= \frac{e^{-s}(2\,e^{s}-1)}{8\,e^{3s}-4\,e^{2s}-2\,e^{s}-1}= 1- 9 s +\frac{137}{2}s^2 - \frac{3129}{3!} s^3 +\ldots\\
  G_B&= \frac{e^{-s}(2\,e^{s}-1)^2}{8\,e^{3s}-4\,e^{2s}-2\,e^{s}-1} =1- 7 s +\frac{103}{2}s^2 - \frac{2359}{3!} s^3 +\ldots\\
  G_C&= \frac{e^{-s}}{8\,e^{3s}-4\,e^{2s}-2\,e^{s}-1} =1- 11 s +\frac{179}{2}s^2 - \frac{4139}{3!} s^3 +\ldots\,,
\end{split}
\end{align}
from which $T_{HTH} = \frac{1}{2}(A+C) = 10$,
$T^2_{HTH} = 2\times \frac{1}{2}(A_2+C_2) = 158$, and
$T^3_{HTH} = 6\times \frac{1}{2}(A_3+C_3) = 3634$, etc.

 \end{document}